# Towards high-throughput superconductor discovery via machine learning

S. R. Xie,[1,2] Y. Quan,[1,2] Ajinkya Hire,[1,2] Laura Fanfarillo,[3,4] G. R. Stewart,[3] J. J. Hamlin,[3] R. G. Hennig,[1,2] and P. J. Hirschfeld[3]

[1]Department of Materials Science and Engineering, University of Florida
[2]Quantum Theory Project, University of Florida
[3]Department of Physics, University of Florida, Gainesville, Florida 32611, USA
[4]Scuola Internazionale Superiore di Studi Avanzati (SISSA) and CNR-IOM, Via Bonomea 265, 34136 Trieste, Italy

**Status:** Even though superconductivity has been studied intensively for more than a century, the vast majority of superconductivity research today is carried out in nearly the same manner as decades ago. That is, each study tends to focus on only a single material or small subset of materials, and discoveries are made more or less serendipitously. Recent increases in computing power, novel machine learning algorithms, and improved experimental capabilities offer new opportunities to revolutionize superconductor discovery. These will enable the rapid prediction of structures and properties of novel materials in an automated, high-throughput fashion and the efficient experimental testing of these predictions.

High-throughput approaches to materials discovery have been successful in situations where a well-defined figure of merit depends directly on simple descriptors available in large databases, e.g., thermoelectric coefficients [1]. Unlike such quantities, the critical temperature $T_c$ of a superconductor depends sensitively on several derived quantities of the coupled electron and phonon systems that are not always well known. Even within the approximate Eliashberg theory of electron-phonon superconductivity, for example, the material-specific Eliashberg function $\alpha^2 F(k,k',\omega)$ is in principle a function of energy and momentum. While $\alpha^2 F$ can be calculated with increasingly high accuracy, the process is still computationally expensive and therefore unsuitable for high-throughput approaches. At the same time, machine learning techniques have improved dramatically, so that one can imagine learning $T_c$ from a simpler discrete set of density functional theory (DFT) based descriptors that parametrize the Eliashberg function and even learn corrections to the Eliashberg theory. At present, Eliashberg codes are quite sophisticated but do not, in general, include corrections due to spin fluctuations from $d$- and $f$-electrons, which can suppress conventional superconductivity or enhance unconventional superconductivity. The recent discovery of near-room temperature superconductivity in high-pressure phases of hydrides by following computational structure predictions illustrates the need for machine learning approaches to accelerate the exploration of the energy landscape of multinary materials. Combining these various concepts, machine learning methods have the potential to improve the prediction of new superconducting materials dramatically.

For decades, the Allen-Dynes (AD) formula, an expression for the $T_c$ of an electron-phonon mediated superconductor, expressed in terms of the moments of the Eliashberg function $\alpha^2 F$ that can be extracted from tunneling experiments, has played an essential role in guiding the search for new superconductors [2]. The AD formula was based on solutions to the Eliashberg equations for a few simple models and materials and is known to deviate strongly from empirical $T_c$'s for high-temperature superconductors, e.g., hydrides. In a proof of principle approach, we used the SISSO analytical machine learning approach to improve on the AD formula, training on the tiny AD dataset of 29 superconducting materials and testing with newer superconducting materials [3]. Clearly, a reliable approach requires a much larger database to learn on, as discussed below.

The search for phonon-mediated high-temperature superconductors rests on a simple principle, i.e., maximizing the electron-phonon coupling strength and the phonon frequencies. Ashcroft proposed [4] dense metallic hydrogen as a candidate for high-temperature superconductivity by noting its high Debye temperature and moderate electron-phonon coupling strength. Nearly half a century after Ashcroft's initial prediction, the hydrogen-rich $H_3S$, rather than pure hydrogen, was found to superconduct at around 200 K under megabar pressures [5].

Machine learning has the potential to accelerate the discovery of novel superconductors, like $H_3S$, by reducing the computational cost of obtaining $T_c$ from first-principles calculations and narrowing down the list of candidate materials. Using data from the Supercon database, Stanev *et al.* carried out machine learning studies to establish connections between the superconducting transition temperatures and atomic properties. They discovered several properties that correlate with the superconducting transition temperatures and predicted a host of compounds as possible high-temperature superconductors [6]. Hutcheon *et al.* constructed neural network models to predict superconductors with high $T_c$ under lower pressure than required for superhydrides thus far [7]. Discovering novel materials by machine learning methods, both with and without coupling to first-principles calculations, is progressing rapidly.

For a given material's system, once thermodynamically stable and metastable crystal structures at various conditions are known, their physical properties, including superconductivity, can be estimated and used to guide the experimental investigations of these materials. Many techniques and codes are available for predicting potentially stable phases and structural transitions, such as the semi-local methods of minima hopping, basin hopping, and simulated annealing and the global methods of genetic algorithms and particle swarm optimization. Chemical intuition, structure-chemistry correlations, and data mining of open-access materials databases can also be used to provide starting structures to these methods. In our work, we have utilized the Genetic Algorithm for Structure and Phase Prediction (GASP) package [8] for exploring possible structural transitions to clathrate-like structures at high pressures in materials containing light elements. While searching for potential stable crystal structures at high pressures, it is essential to also consider the stability of the competing phases that might destabilize the stoichiometry of the material of interest. To this end, the GASP package can perform grand canonical searches, meaning it can search for stable crystal structures over the entire composition range of interest and generate a high-pressure convex hull diagram.

While the discovery of the high-pressure superconducting hydrides has revolutionized the field of electron-phonon superconductivity scientifically, it is of little practical utility unless methods can be found to engineer materials at high pressure that remain metastable under ambient conditions. One approach has been to search for hydrides that are superconductors at lower pressures [9]. However, the route to ambient pressure high-$T_c$ materials along this path is far from clear. One promising approach to ambient pressure superconductors that may be used to make large scale films is non-equilibrium growth of metastable materials on solid substrates by methods such as molecular beam epitaxy or chemical vapor deposition, potentially assisted by laser heating. Employing high-energy pre-mixed amorphous starting materials to reach structures and compositions that are inaccessible by just compressing ambient-pressure crystal structures could provide synthesis routes to novel high-pressure phases that remain metastable at ambient conditions. Theory can assist by predicting and characterizing metastable crystal structures to guide the choice of atomic constituents and estimate barriers to the equilibrium phase.

**Current and Future Challenges**

Machine learning and high-throughput methods work best when applied to large datasets. Given the improvement of first-principles based predictions of $T_c$ for electron-phonon based superconductors in recent years, one could imagine constructing a database to enable machine learning of higher-$T_c$ superconductors by calculating $\alpha^2 F$ and associated moments, e.g., for all 16,414 materials in the SuperCon database. However, such a program is prohibitively expensive since each $\alpha^2 F$ calculation typically requires O(100-1000) core-hours of computing time even for elemental superconductors with a modern package such as Quantum Espresso. Purely electronic descriptors like the Fermi level density of states, $N(0)$, are much cheaper and can enable extensions of machine learning approaches such as that attempted by Stanev et al. [6]. The challenge, in this case, is to identify which purely electronic descriptors are important for high-$T_c$. We are constructing a high-throughput scheme along these lines.

Any high-throughput approach for novel superconductors requires robust structure predictions. Genetic algorithm searches and other methods coupled to first-principles relaxations are often limited in their prediction capabilities by the number of atoms in the supercell. As the number of atoms increases, the number of local minima in the energy landscape rises exponentially and the computational cost of relaxations grows polynomially. The room-temperature superconductors of tomorrow might potentially have large unit cells and may contain more than 3 elements. The crystal structure prediction of such superconductors can be accelerated by utilizing machine-learned surrogate models of the energy landscape that are trained on small structures.

Many of the techniques used until now in search of high-temperature electron-phonon superconducting materials may be applied to the analysis of unconventional superconductors, where Cooper pairs form due to the exchange of electronic rather than phononic excitations. Examples of unconventional superconducting materials are cuprates, iron-based superconductors, and heavy fermions. Unlike electron-phonon systems, however, a unified theory of superconductivity that could explain the variety of phenomenology shown by the various families of unconventional superconductors is not yet available. Approximate, often successful but uncontrolled methods formulated based, for example, on spin fluctuation-mediated superconductivity [9] have not been convincingly coupled to first-principles electronic structure calculations, and therefore DFT-based descriptors important for unconventional superconductivity are less well understood.

**Advances in Science and Technology to Meet Challenges**

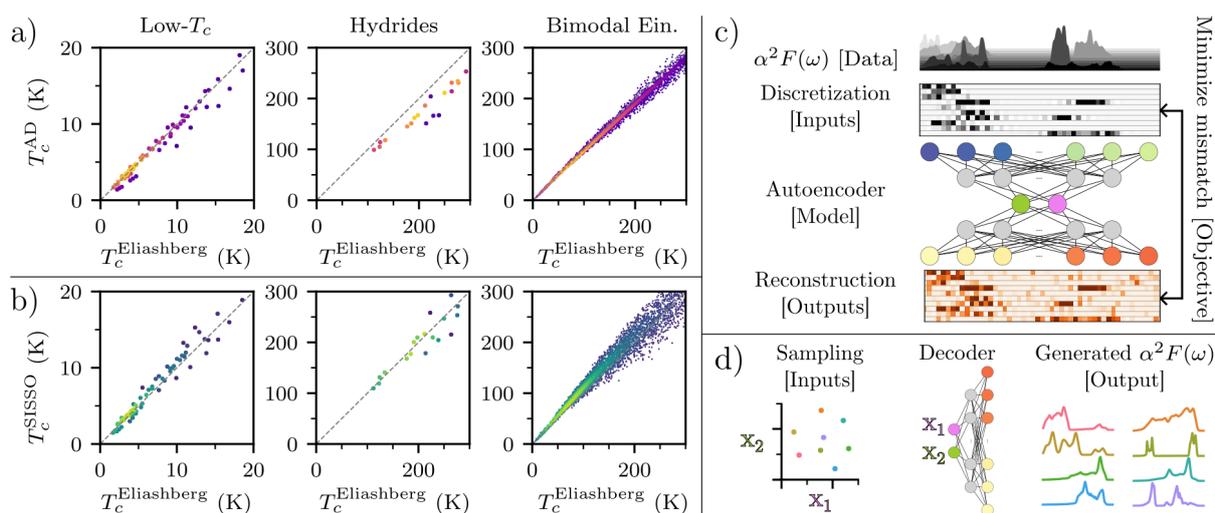

**Figure 1.** Performance comparison between the Allen-Dynes equation [2] and a model fit with symbolic regression alongside a proposed workflow for data augmentation. a) Allen-Dynes performance on low-$T_C$, hydrides, and bimodal Einstein data. Plots are shaded by the log-density of points. b) Performance of symbolic regression machine learning model. c) Workflow for training an autoencoder neural network to learn an efficient compression of $\alpha^2F$. The mismatch between the original input and the reconstructed output, after compression and decompression, is minimized iteratively. d) The latter half of the trained autoencoder, known as the decoder, generates new $\alpha^2F(\omega)$ spectra by sampling the learned distribution.

Recently, we extended the approach to study a much larger sample of $\alpha^2F$'s generated from a) EPW electron-phonon calculations for 50 real materials and 5,000 artificial $\alpha^2F$'s of bimodal form. The solution of the isotropic Eliashberg equation, given $\alpha^2F$, is relatively inexpensive. Using the resulting dataset, which is two orders of magnitude larger than that of the previous work [3], we obtained a predictive model with improved performance over the AD formula for systems with higher $T_c$. Figure 1a illustrates the performance of the AD formula on low-Tc materials such as elemental metals, hydrides such as $LaH_{10}$ and $H_3S$, and the bimodal einsteins used to augment the training dataset. The systematic underprediction of $T_C$ with the AD formula is absent from predictions made with a new machine-learned equation, shown in Figure 1b. Moreover, we were able to determine that deviations of AD from Eliashberg theory, for large λ systems, arose primarily from the inequivalence of the $\alpha^2F$ moments $\omega_2$ and $\omega_{log}$.

Based on the success of augmenting the data using model systems based on multimodal einsteins, it is clear that generative models are another promising avenue for improved machine-learning of $T_C$. Figure 1c and 1d illustrate one method of generating $\alpha^2F$ spectra using an autoencoder neural network. Training data, including known examples of $\alpha^2F$, are first discretized through binning or basis set expansion. The autoencoder learns an efficient compression of the data into a lower-dimensional "latent" space, shown here as dimensions $x_1$ and $x_2$, by iteratively minimizing the error between the input and the reconstructed output. The trained decoder, comprised of the layers after and including the autoencoder bottleneck, can then transform any sampled point in the latent space into a $\alpha^2F$. Coupled with the Eliashberg equations, this approach augments the dataset with arbitrarily-many training samples that are qualitatively realistic compared to model systems.

For unconventional superconductors, a more promising approach is to use empirical knowledge about the normal state of systems that support unconventional superconductivity. For example, unconventional superconductivity is often found near phase transitions where magnetism disappears via doping or applied pressure. Such "soft" magnetic states can be identified by high-throughput DFT calculations of magnetic candidate materials (mostly metallic compounds in materials database with transition metal ions), calculating how magnetism responds to applied pressure. Promising materials where magnetization decreases rapidly can then be synthesized and studied further by doping and/or pressure. A more theoretically guided approach is based on a weak-coupling scenario in which spin-excitations provide the effective attractive interactions between electrons. In this case, our strategy is to look for the optimal conditions to find a large spin susceptibility peaked at a particular momentum that could reflect in a strong pairing at this wave vector. Nesting wave vectors on Fermi surfaces are often proxies for such peaks and can also be searched for in high-throughput schemes.

**Concluding Remarks**

We have outlined an approach to superconductor discovery that seeks to identify the electronic properties of a material that are essential for high $T_c$ via machine learning techniques. One thrust is to develop improved equations for $T_c$ of the Allen-Dynes type capable of accounting for large-$\lambda$ materials like high-pressure superconducting hydrides. We discussed the challenges of creating sufficiently large databases of Eliashberg $\alpha^2F$'s using expensive electron-phonon calculations for real materials and addressed them by the creation of artificial $\alpha^2F$'s with modeling of simple physically motivated form, and the use of autoencoders, both equally good for learning Eliashberg theory. We also discussed future directions, including structure prediction utilizing surrogate machine learning models of energy landscapes to inform studies of metastable materials synthesized under high pressure, and extension of current methods to unconventional superconductors.

**Acknowledgments**

*The work was supported by the US Department of Energy Basic Energy Sciences under Contract No. DE-SC-0020385.*

**References**

[1.] S. Curtarolo, Gus L. W. Hart, M. Buongiorno Nardelli, N. Mingo, S. Sanvito and O. Levy, ''The high-throughput highway to computational materials design", Nat. Mat. 12, 191 (2013).

[2.] P. B. Allen and R. C. Dynes, "Transition temperature of strong-coupled superconductors reanalyzed", Phys. Rev. B 12, 905 (1975).

[3.] S. R. Xie, G. R. Stewart, J. J. Hamlin, P. J. Hirschfeld, and R. G. Hennig, "Functional Form of the Superconducting Critical Temperature from Machine Learning", Phys. Rev. B 100, 174513 (2019).

[4.] N. W. Ashcroft, "Metallic Hydrogen: A High-Temperature Superconductor?", Phys. Rev. Lett. 21, 1748 (1968).

[5.] A. Drozdov, M. Eremets, I. Troyan, *et al.,* "Conventional superconductivity at 203 kelvin at high pressures in the sulfur hydride system". *Nature* **525,** 73 (2015).

[6.] V. Stanev, C. Oses, A. Gillad Kusne, E. Rodriguez, J. Paglione, S. Curatolo, and I. Takeuchi, "Machine learning modeling of superconducting critical temperature", Nat. Comp. Mat. 4, 29 (2018).

[7. ] Michael J. Hutcheon, Alice M. Shipley, and Richard J. Needs, "Predicting novel superconducting hydrides using machine learning approaches", Phys. Rev. B 101, 144505 (2020).

[8.] W. W. Tipton and R. G. Hennig, "A grand canonical genetic algorithm for the prediction of multicomponent phase diagrams and testing of empirical potentials," J. Phys. Condens. Matter, vol. 25, no. 49, p. 495401, 2013.

[9.] D. J. Scalapino*, "*A common thread: The pairing interaction for unconventional superconductors*",* Rev. Mod. Phys. 84, 1383 (2012).